\documentstyle[aps,prl,twocolumn,epsf]{revtex}
\begin{document}
\input{psfig.sty}
\draft
\newcommand{\h}{\mbox{$\frac{1}{2}$}}
\newcommand{\th}{\mbox{$\frac{3}{2}$}}
\newcommand{\bd}{\mbox{${\bf d}$}}
\newcommand{\p}{\mbox{$^{\prime}$}}

\twocolumn[\hsize\textwidth\columnwidth\hsize\csname
@twocolumnfalse\endcsname

\title{Anisotropic spin freezing in the S=1/2 zigzag ladder compound $\rm SrCuO_2$}

\author{I.~A.~Zaliznyak$^{1,2,}$\cite{Kapitza}, C.~Broholm$^{1,2}$, M.Kibune$^3$, M. Nohara$^3$, and H.Takagi$^3$}
\address{
$^1$Department of Physics and Astronomy, The Johns Hopkins University, Baltimore, Maryland 21218\\
$^2$National Institute of Standards and Technology, Gaithersburg, Maryland 20899\\
$^3$Graduate School of Frontier Sciences and Institute for Solid State Physics,
University of Tokyo, Hongo, Tokyo 113-8656, Japan and CREST-JST}

\date{\today}
\maketitle

\begin{abstract}

Using magnetic neutron scattering we characterize an unusual low temperature phase in orthorhombic SrCuO$_2$. The material contains zigzag spin ladders formed by pairs of S=1/2 chains ($J=180$ meV) coupled through a weak frustrated interaction ($|J^{\prime}|\lesssim 0.1J$). For $T<T_{c1}=5.0(4)$ K an elastic peak develops in a gapless magnetic excitation spectrum indicating spin freezing on a time scale $\delta t>\hbar /\delta E=2\cdot 10^{-10}$ s. While the frozen state has  long range commensurate antiferromagnetic order along the chains ($\xi_c>200c$) and a substantial correlation length ($\xi_a=60(25) a$) perpendicular to the zigzag plane, the correlation length is only $\xi_b=2.2(3) b$ in the direction of the frustrated interaction. We argue that slow dynamics of stripe-like cooperative magnetic defects in tetragonal $\bf a-c$ planes yield this anisotropic frozen state.

\end{abstract}


]
Reduced dimensionality\cite{Steiner,DeJongh}, geometrical
frustration\cite{Ramirez}, and band formation\cite{Aeppli}
can suppress the critical temperature of  magnets far below a
cooperative energy scale such as the Curie-Weiss
temperature. The resulting low temperature ordered phases have
unconventional features that challenge conventional theories of
magnetism. They include strongly reduced sublattice magnetization,
abnormal sensitivity to low levels of disorder, and mean-field like
critical behavior.

Quasi-one-dimensional magnetic dielectrics are excellent model systems
in which to explore weakly ordered phases because a quantitative 
connection between the dynamic properties of the one-dimensional 
units and critical phenomena in the coupled system can be 
established \cite{Scalapino75,Affleck,Schulz96}. 
Haldane spin-1 chains and other spin systems with a gap have a 
critical value for the inter-chain coupling $zJ_{\perp}\approx\Delta$ 
below which they remain disordered down to $T=0$. 
Spin-1/2 chains on the other hand, are predicted to order 
even for vanishingly small non-frustrated inter-chain interactions. 
This is consistent with the low temperature long range order among 
weakly coupled spin-1/2 chains found in $\rm Sr_2CuO_3$ and 
$\rm Ca_2CuO_3$\cite{Kojima97}.

In this letter we describe a different low temperature phase in 
closely related $\rm SrCuO_2$, which contains linear spin-1/2 chains 
assembled pairwise in an array of weakly interacting zigzag ladders 
\cite{Ishida,Matsuda,Teske,Matsuda2,Motoyama,Tanaka,Ohta}. 
Though weak static sublattice magnetization does develop for 
$T< 5.0(4)$K $\approx 2.8\cdot 10^{-3}J/k_B$, three dimensional long 
range order is absent for $T\gtrsim 10^{-4}J/k_B$. 
Specifically, we find a low temperature correlation length of 
only two lattice spacings along the direction of frustrated 
intra zigzag ladder interactions. 
We argue that slow dynamics of stripe-like defects 
in quasi-two-dimensional antiferromagnetic layers prevent the system from developing 
long range order. 

Zig-zag ladders in SrCuO$_2$ are built from corner-sharing Cu-O 
chains with an exchange constant $J\approx 181(17) $ meV 
\cite{Motoyama} stacked pairwise in edge-sharing geometry. 
The frustrating interaction between chains proceeds through 
$\approx 87.7^\circ$ Cu-O-Cu bonds \cite{Matsuda,Rice} and is 
expected to be weak and ferromagnetic $|J'|\lesssim 0.1J$. 
Field theory\cite{Haldane1,Allen,White} and numeric simulations 
\cite{White,Chitra,Normand} for a pair of chains coupled like this 
predict that weak antiferromagnetic inter-chain coupling 
induces incommensurate correlations and a gap $\Delta\sim
J\exp(-\alpha J/J')$ while the zigzag chain should remain gapless for
ferromagnetic $J'$ \cite{Allen,White}.

SrCuO$_2$ is centered orthorhombic (space group $Cmcm\equiv D^{17}_{2h}$) with lattice 
parameters $a=3.556(2)\AA$, $b=16.27(4)\AA$, $c=3.904(2)\AA$
\cite{Teske}. We index wave
vector transfer in the corresponding simple orthorhombic reciprocal
lattice. Nuclear Bragg reflections $(h,k,l)$ are allowed for even 
$h+k$ and even $l$ when $k=0$. 
While there is no direct information about the magnitude of 
interactions between zigzag ladders, the structural features that 
determine them are similar to those in $\rm Sr_2CuO_3$. 
Application of quantum Chain Mean Field (CMF) theory \cite{Schulz96} 
to $\rm Sr_2CuO_3$ yields an estimate for inter-chain interactions of 
$\overline{J_{\perp}}\approx k_BT_N/(1.28\sqrt{\ln (5.8 J/k_BT_N)})
\approx 0.13$ meV, while according to classical \cite{Scalapino75} CMF 
$\overline{J_{\perp}}=3(k_BT_N)^2/(8JS^2(S+1)^2)\approx 8\cdot 
10^{-4}$ meV.

The neutron scattering experiments were performed at the NIST Center
for Neutron Research using cold and thermal neutron triple axis
spectrometers.
PG(002) reflections were used for monochromator and analyzer, 
supplemented by Be and PG filters. Our sample is a 
cylindrical rod ($\ell\approx 46$ mm, $D\approx 5$ mm, and 
$m= 3.875(5)$ g) grown by the Traveling Solvent
Floating Zone (TSFZ) technique.  Rocking curves about the $\bf b$ axis showed
two roughly equal intensity peaks separated by $\approx 0.5^\circ$ 
and each with a Full Width at Half Maximum (FWHM) $\approx 0.25^\circ$.
Experiments were performed with wave-vector transfer in the $(h,0,l)$ 
and $(h,k,h)$  reciprocal lattice planes. 
We used incoherent scattering from a vanadium

\vspace{0.15in}
\noindent
\parbox[b]{3.4in}{
\psfig{file=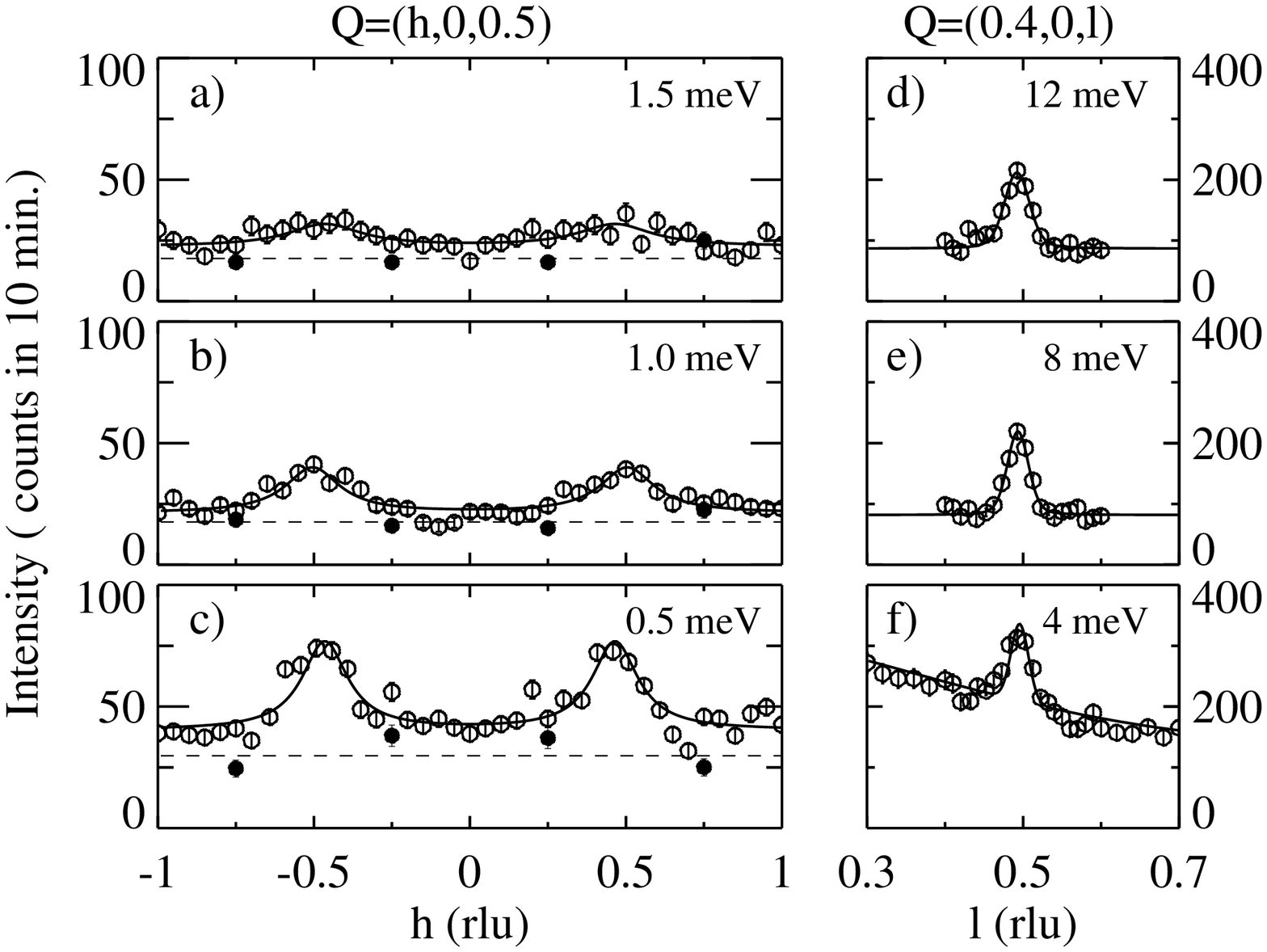,width=3in}
\vspace{0.3in}
{FIG.~1. \small
Wave-vector dependence of  inelastic magnetic scattering from
SrCuO$_2$.  (a)-(c) show data along $\bf a^*$ which is perpendicular to
the zigzag ladders while (d)-(f) show data along the chains.  (a)-(c)
were collected on SPINS at $T=0.35$K with $E_f=5.1$~meV and
collimations $80^{\prime}-80^{\prime}-240^{\prime}$.  (d)-(f) come from
BT2 at $T=12.5(5)$K with $E_f=14.7$~meV and collimations
$60^{\prime}-60^{\prime}-80^{\prime}-120^{\prime}$. Increasing
background in (f) comes from the direct beam. }}

\vspace{0.05in}

\noindent
rod in the same geometry as our $\rm SrCuO_2$ sample to normalize the
magnetic scattering cross section.

The defining characteristic of a quasi-one-dimensional spin system is
an anisotropic dynamic correlation volume, which can be probed 
by inelastic magnetic neutron scattering. 
Fig. 1 shows constant energy scans along two perpendicular
directions in the $(h,0,l)$ plane for energy transfer 0.5
meV$<\hbar\omega< 14$ meV. Scans along the chain direction (Fig. 1
(d)-(f)) reveal a resolution limited peak centered at $l=\h$. The data
yield an upper limit $\delta q\lesssim 0.01{\bf c^*}$ on any 
intrinsic FWHM of the peak. For comparison, the FWHM of the des 
Cloizeaux-Pearson continuum for each antiferromagnetic spin-1/2
chain at $\hbar\omega=4$ meV is predicted to be $\delta q=
(2\hbar\omega/\pi^2J)c^*=4.5\cdot 10^{-3}$ and hence it is unresolved 
in our measurement. 
Fig. 1 (a)-(c) show scans along the $\bf a$ direction which is normal 
to the plane of the zigzag ladders. There we find peaks for 
$|h|\approx \h$ which span the better part of the Brillouin zone and 
this provides evidence for short range dynamic antiferromagnetic 
correlations perpendicular to zigzag ladders.
The correlation anisotropy between the two directions probed is $\delta
q_a/\delta q_c>10a^*/c^*$. While our measurements do not yield an
energy scale associated with dispersion along $\bf c^*$ we found that
modulation along $\bf a^*$ exists only for 
$\hbar\omega \lesssim 2$ meV.

Susceptibility, heat capacity, and $\mu SR$ measurements indicate that
static magnetic order develops below $T \sim 2$K in SrCuO$_2$ 
\cite{Matsuda2}. Figure 2 shows the corresponding elastic magnetic 
neutron scattering, which we found at ${\bf Q}=\tau\pm{\bf Q}_m$ where
$\tau$ is a reciprocal lattice vector, 
${\bf Q}_m=(\h+\epsilon,k,\h)$, $\epsilon=0.006(1)$, and $k$ is an 
integer. However, rather than being
concentrated in a   resolution limited 

\noindent
\parbox[b]{3.4in}{
\psfig{file=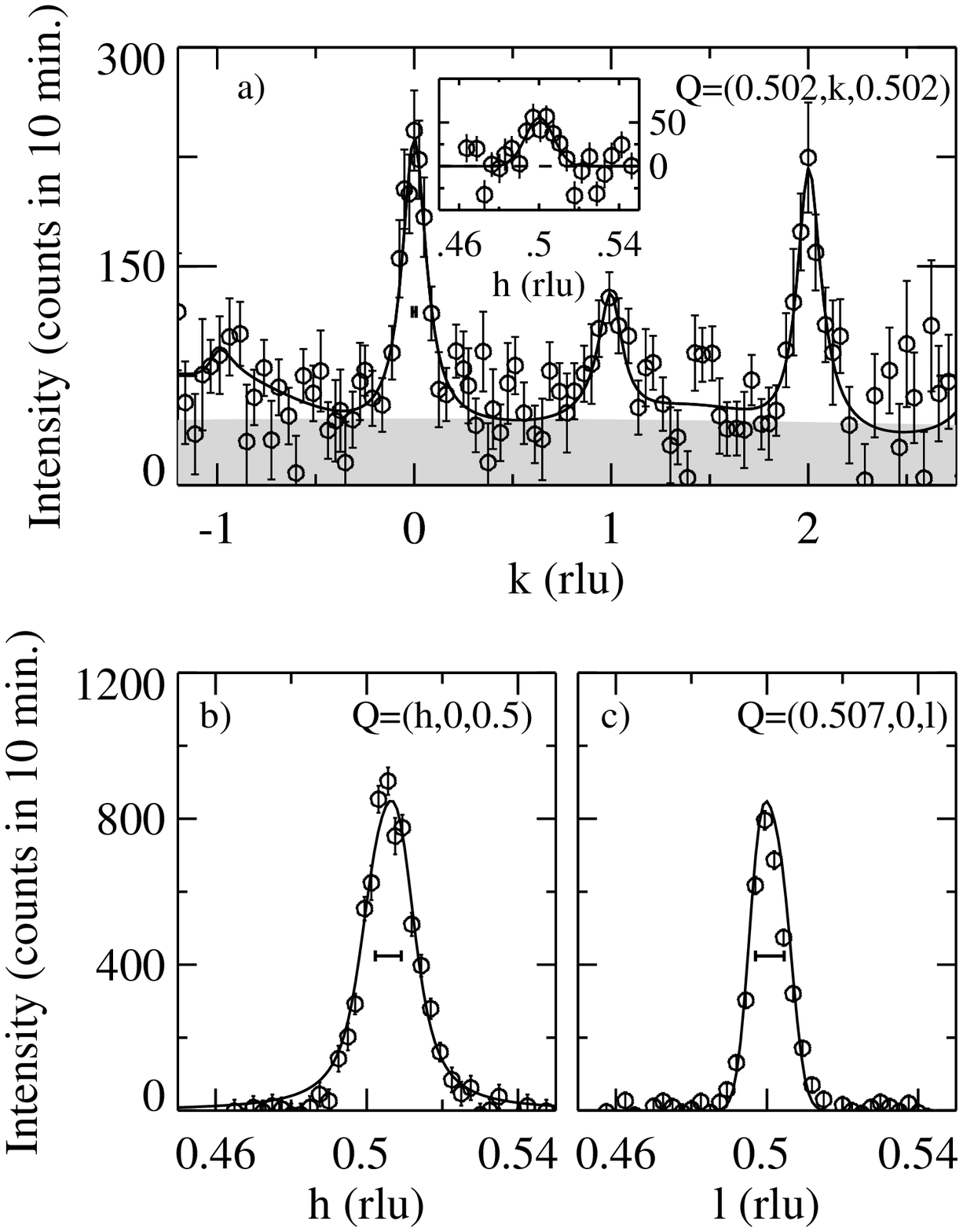,width=3.5in}
{FIG.~2. \small
(a)-(c) wave-vector dependence of the net elastic magnetic intensity 
around ${\bf Q}\approx (0.5,0,0.5)$ at T=0.35K along $\bf b^*$, 
$\bf a^*$ and $\bf c^*$ directions respectively. The background was 
measured at T=8K. 
Horizontal bars show calculated FWHM of the instrument resolution 
function. 
(a) BT2 and BT9 measurements with $E_f=14.7$~meV and collimations 
$60'-40'-40'-120'$. The shaded area shows $k$-independent 
``magnetic rod'' intensity. The solid line shows Eq. (1) with 
optimized Fourier coefficients and multiplied by the squared 
magnetic form factor for Cu$^{2+}$.
Insert shows scan across the rod at $k=0.35$. (b),~(c) SPINS data with 
$E_f=3.7$~meV and collimations $80'-40'-240'$. Solid lines are fits to 
resolution convoluted Lorentzians. 
}}

\vspace{0.05in}

\noindent
Bragg peak, the signal has a characteristic distribution in reciprocal space, which
shows that our sample does not have conventional three dimensional
long-range magnetic order. The peak is sharpest in the $\bf c^*$
direction, wider and incommensurate in the $\bf a^*$ direction, and
very broad along $\bf b^*$. Fig. 2 (a) and its inset also reveal
a rod of elastic magnetic scattering extending along $\bf b^*$.
Based on this information the frozen magnetic structure can be 
described as follows. Spins in each tetragonal $\bf a-c$ plane are 
aligned antiferromagnetically with a superposed long 
wavelength modulation or defect structure in the $\bf a$ direction. 
Such planes are stacked with a correlation length of only a few times 
the lattice spacing along $\bf b$.

By fitting data in Fig. 2 to Lorentzians duly convoluted with the
instrumental resolution function, the magnetic correlation lengths in
the $\bf a-c$ plane were determined to be $\xi_a=60(25)a$ and
$\xi_c\gtrsim 200 c$ . The peak position refined to ${\bf
Q}_m=(0.506(1),0.00(1),0.500(1))$. Note that the incommensurability
along $\bf a^*$ was reproduced in several independent experiments.
Moreover, we found an equivalent magnetic satellite from $\tau=(202)$ 
at ${\bf Q}=(1.494(3),0.00(1),1.500(1))\approx\tau-{\bf Q}_m$.

The extremely short correlation length along $\bf b$ suggests analysis
of scans along this direction in terms of a Fourier series:
\begin{equation}
{\bar S}^{\alpha\alpha}({\bf Q})=\frac{\langle S\rangle ^2}{3}
(1+\frac{1}{N_{\parallel}}\sum_{j\neq j^{\prime}}
{\cal C}_{j,j^{\prime}}\cos{\bf Q}\cdot (\bd_j-\bd_{j^{\prime}}) ),
\end{equation}
where $j,j^{\prime}$ index consecutive $\bf a-c$ spin planes and
${\cal C}_{j,j^{\prime}}=(1/N_{\parallel})\sum_m
\langle S_{\bd_{j+2m}}\rangle\langle S
_{\bd_{j^{\prime}+2m}}\rangle/\langle S\rangle ^2$.
For simplicity we have assumed that the frozen spin configuration is
isotropic in spin space and an exponential decay of correlations 
between planes separated by distances of $2b$ and beyond. 
The line through the data in Fig. 2 shows the result of this fit. 
The correlation length extracted is $\xi_b = 2.2(3) b$. The 
correlation parameters ${\cal C}_{jj^{\prime}}$ are listed in table 1.  

$\rm SrCuO_2$ is built from copper bilayers.
Equivalent spins within a bilayer are displaced by
$\bd_1-\bd_0=(0\delta \h)$ to form zigzag ladders. Bilayers are stacked
in registry -ABABAB- where neighboring A and B type bilayers are
separated by $\bd_2-\bd_0=(\h\h 0)$. Table 1 shows that
despite the proximity of planes that make up a bilayer, 
correlations between them are weak. This is experimental evidence for 
an effective decoupling between zigzag ladder rungs. 
Looking beyond a bilayer we see antiferromagnetic correlations 
between spins in shifted bilayers A and B (${\cal C}_{0\bar{1}}, 
{\cal C}_{03},  {\cal C}_{0\bar{5}}, {\cal C}_{07}<0$) and 
ferromagnetic correlations between spins in bilayers separated by 
multiples of $\bf b$  (${\cal C}_{0\bar{3}}, {\cal
C}_{0\pm 4}, {\cal C}_{05}, {\cal C}_{0\bar{7}}>0$).  Though they are
separated by half the distance, nearest neighbor A-B correlations are
significantly weaker than A-A correlations and this is evidence for
frustrated inter-bilayer interactions. From the fit we also obtain a
reliable value for the frozen moment at $T=0.3$ K which turns out to be
only a minute fraction of the full moment:  $g <S>=0.033(7)$ per Cu.

The temperature dependence of the squared frozen moment and the
in-plane correlation parameters are shown in Fig.~3.  The frozen
staggered magnetization appears below $T_{c1}=5.0(4)$K and initially
increases in proportion to $(T_{c1}-T)^{2\beta}$ where 
$\beta=0.46(12)$. Below an inflection point at $T_{c2}=1.5(3)$K, 
$\langle S\rangle^2$ increases faster until saturating below 
$T\sim 0.5$K. Careful inspection of specific heat data\cite{Matsuda2} 
for $\rm SrCuO_2$ reveal anomalies close to both $T_{c1}$ and 
$T_{c2}$. While the peak position along the chain, $l_0$, is 
temperature independent, both the incommensurability and the Half 
Width at Half Maximum (HWHM), $\kappa_h=a/(\pi\xi_a)$ along $\bf a^*$, 
decrease by approximately a factor 2 below $ T_{c2}$. It is useful to 
compare our results to those for $\rm Sr_2CuO_3$ \cite{Kojima97}. 
The critical temperature ($T_N=5$ K) and intra-chain exchange constant
($J=220$ meV) are similar for the two materials. Nonetheless, the 
moment in $\rm SrCuO_2$ is almost twice smaller than in 
$\rm Sr_2CuO_3$.

\noindent
\parbox[b]{3.4in}{
\psfig{file=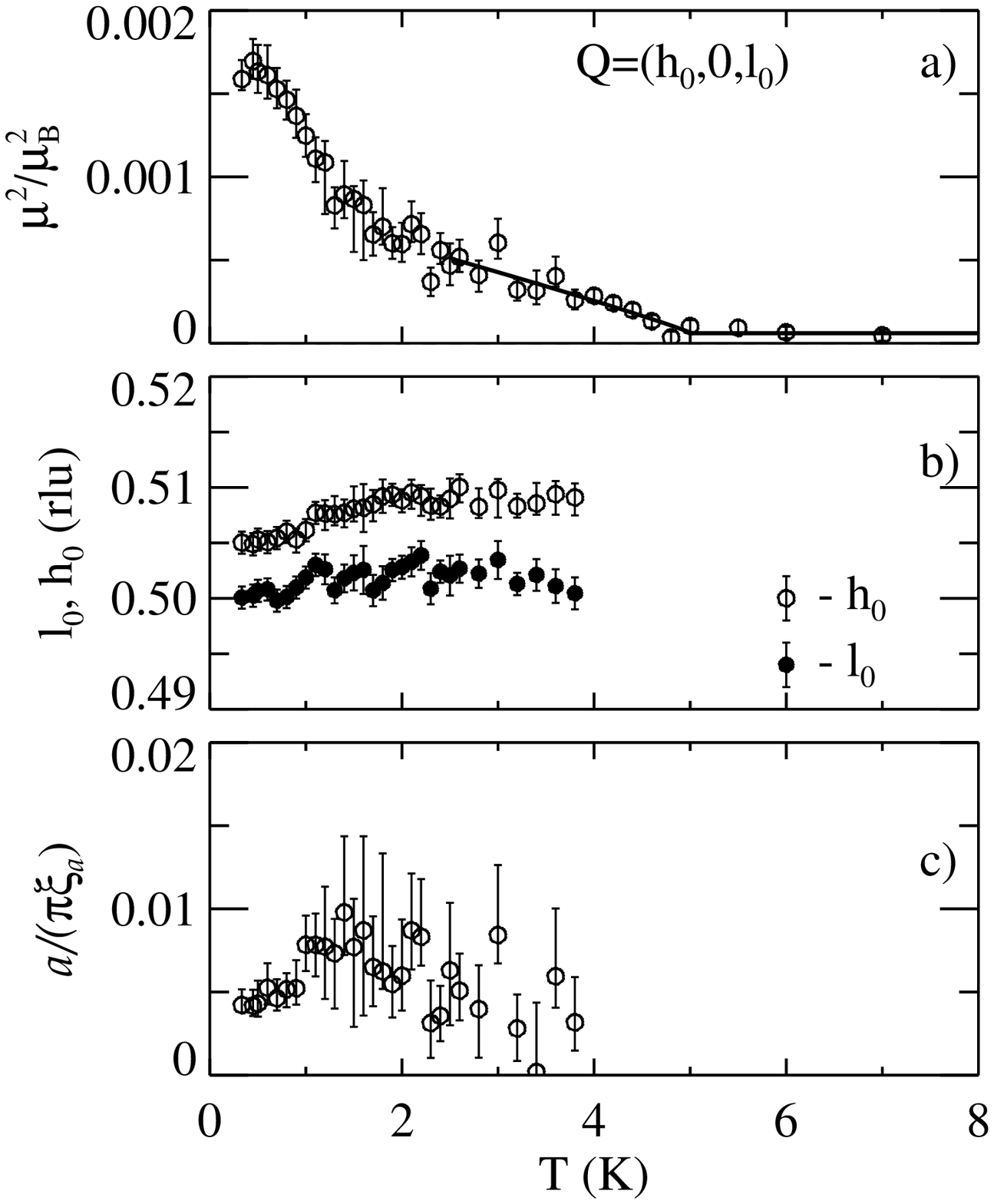,width=3.5in}
\vspace{-0.2in}
{FIG.~3. \small
Temperature dependencies of (a) the frozen sublattice magnetization
squared, (b) the in plane wave vector (${\bf Q}=(h_0.0.l_0)$), and (c)
the inverse correlation length along $\bf a$. Data derived from $h$ and
$l$ scans through the $(\h+\epsilon 0 \h)$ elastic magnetic peak on
SPINS using $E_f=4.47$meV and collimations $80\p -80\p -240\p $. The frozen
moment in (a) is within error bars of our global average.
}}

\vspace{0.05in}


\noindent
There is also a qualitative distinction between ordering in these 
materials; the frozen phase in $\rm SrCuO_2$ is incommensurate along 
$\bf a$, while $\rm Sr_2CuO_3$ has commensurate order\cite{Kojima97}. 
Magnetic disorder along $\bf b$ is also unique
to  $\rm SrCuO_2$. While this feature may be connected to quenched
disorder, possibly from oxygen non-stoichiometry in our sample, the
magnetic correlation length along $\bf b$ is easily an order of
magnitude less than the defect spacing.  $\rm SrCuO_2$ therefore falls
in the category of magnets which are either intrinsically disordered or
display an extreme sensitivity to disorder. We shall argue that 
highly frustrated intra- and inter-zigzag ladder interactions are 
central to understanding these differences, and the anisotropic frozen 
state in general.

The main 
contribution to inter-zigzag ladder interactions along {\bf a} 
comes from weak superexchange along Cu-O-O-Cu 
paths with two $\approx 90^\circ$ bonds. 
For each spin there are 12 such equivalent Cu-O-O-Cu paths to 
neighbors in the same {\bf a}--{\bf c} plane; four to each nearest neighbor 
along {\bf a}, denote the corresponding exchange constant $J\p_a>0$, and one
to each next nearest neighbor in diagonal $(1,0,\pm 1)$ directions, $J\p\p_a>0$. 
As a consequence we expect that $J\p\p_a\sim J\p_a/4$ corresponding to a 
highly frustrated point for this lattice. 
Similar interactions exist in the 
$(1,\delta,\pm\frac{1}{2})$ directions between nearest neighbors in 
adjacent $\bf a-c$ planes.
In fact, it is evident that these interactions as well as the inter-bilayer 
interactions all share the 
same frustrating zigzag geometry and all cancel at the mean-field level. In addition there may be other weak  non-frustrated 
interaction  favoring long range order.
Experimental evidence for competing interactions in $\rm SrCuO_2$ 
lies in the incommensurability of the magnetic peaks along the 
$\bf a^*$ direction. We note that 
the shift of the magnetic peak along $\bf a^*$ equals its HWHM not 
only at the lowest $T$ but also as a function of $T\lesssim T_{c2}$. 
This indicates that we are dealing not with a periodic
modulation superimposed on otherwise antiferromagnetic $\bf a-c$ planes
but a highly disordered or even random sequence of {\it frozen} 
stripe defects induced by these competing interactions.

Recent theoretical studies \cite{Affleck,Haldane2} have established
that coupled S=1/2 chain systems can remain disordered at T=0 if 
interchain interactions are sufficiently frustrated.  In the quasi-two-dimensional
case which may be a good first approximation for $\rm SrCuO_2$,
disorder at T=0 could result from {\it instanton} topological defects
created through quantum tunneling \cite{Haldane2}. For {\it
half-integer} or {\it odd} integer spin systems the disordered ground
state is predicted to be degenerate and gapless while {\it even}
integer spin systems should be gapfull.  Our experimental results
support a gapless phase for S=1/2.

At $T>0$ topological disorder is enhanced through thermal creation of 
defects, leading to a reduced correlation length. 
In the coupled S=1/2 chain system we expect defects to be strongly 
anisotropic, taking the form of stripes extending along $\bf c$ between 
bands of phase shifted antiferromagnetic domains. For order to 
develop in a stack of $\bf a-c$ planes, stripe defects in neighboring 
planes must move into registry. However, as mentioned above, the 
residual interactions which favor such ordering are either exceedingly weak or 
cancel at the mean field level.
Thus it seems plausible that pinning and/or intrinsically slow dynamics of
such stripe defects might lead to {\it spin freezing} instead of long range order.

In summary we have found an anisotropic spatially disordered frozen
spin configuration among interacting zigzag spin-1/2 ladders in $\rm
SrCuO_2$. The magnetic correlation length along the short direction of
the zigzag ladder is far less than the impurity spacing. Highly 
frustrated interactions both within the zigzag spin-1/2 ladders and 
between ladders, as well as slow dynamics and pinning of 
order-destroying stripe defects are likely reasons that $\rm SrCuO_2$ behaves differently from the closely related linear 
chain system $\rm Sr_2CuO_3$. Previous theoretical work on coupled 
S=1/2 chains has indicated the possibility of an intrinsic disordered
phase\cite{Affleck,Normand,Haldane2}. Given our data it would be interesting 
to further explore this possibility with competing interactions of 
the specific type found in $\rm SrCuO_2$.

We thank I.~Affleck, G.~Shirane, and A. J. Millis for useful
discussions and R.~Erwin for assistance during the experiments. Work at
JHU was supported by the NSF through DMR-9453362. This work used
instrumentation supported by NIST and the NSF through DMR-9423101.

\begin{table}
\caption{
Correlations between spins in $\bf a-c$ planes
displaced by $\bd_{j^{\prime}}-\bd_0$ 
with respect to each other.
Coordinates are given in units of orthorhombic cell parameters.
$\delta=0.122$\protect\cite{Teske} is the spacing in units of b between
linear spin chains that form a zigzag chain. Error bars define intervals
wherein $\chi^2$ is statistically indistinguishable from its minimum value. ``?" indicates
correlations whose contributions to Eq. (1) cancel in a $(\h k\h )$ scan.   
}
\begin{tabular}{rlrc||rlr}
$j^{\prime}$&$\bd_{j^{\prime}}-\bd_0$ & ${\cal C}_{0,j^{\prime}}$  & &
$j^{\prime}$&$\bd_{j^{\prime}}-\bd_0$ & ${\cal C}_{0,j^{\prime}}$\\
\tableline
$\ 1$&$(0\delta \h)$ & -0.09(6) & &$\ 5$&$(01+\delta\h)$ & 0.09(4)\\
$-1$&$(\h\delta-\h\h)$ & -0.09(5) & &$-5$&$(\h\delta-\th\h)$ & -0.18(4)\\
$\pm2$&$(\h\pm\h 0)$     &   ?      & &$\pm6$&$(\h\pm\th0)$     &    ?\\
$\ 3$&$(\h\h+\delta\h)$ & -0.19(5) & &$\ 7$&$(\h\th+\delta\h)$& -0.11(4)\\
$-3$&$(0\delta-1\h)$   & 0.06(5)  & &$-7$&$(0\delta-2\h)$   & 0.06(4)\\
$\pm 4$&$(0\pm 1 0)$       & 0.23(3)& & & &\\
\end{tabular}
\end{table}

\end{document}